\documentclass[preprint2]{aastex}

\shortauthors{C. Akerlof et al.}
\shorttitle{GRB Prompt Optical Observations}

\begin{document}

\title{Prompt Optical Observations of Gamma-ray Bursts}

\author{Carl Akerlof$^1$, Richard Balsano$^2$, Scott Barthelmy$^3$, 
Jeff Bloch$^2$, Paul Butterworth$^3$, Don Casperson$^2$, 
Tom Cline$^3$, Sandra Fletcher$^2$, Fillippo Frontera$^4$, 
Galen Gisler$^2$, John Heise$^5$, Jack Hills$^2$, Kevin Hurley$^6$, 
Robert Kehoe$^1$, Brian Lee$^{1,9}$, Stuart Marshall$^7$, Tim McKay$^1$, 
Andrew Pawl$^1$, Luigi Piro$^8$, John Szymanski$^2$ and Jim Wren$^2$}

\affil{$^1$University of Michigan, Ann Arbor, MI 48109}
\affil{$^2$Los Alamos National Laboratory, Los Alamos, NM 87545}
\affil{$^3$NASA/Goddard Space Flight Center, Greenbelt, MD 20771}
\affil{$^4$Universit\`{a} degli Studi di Ferrara, Ferrara, Italy}
\affil{$^5$Space Research Organization, Utrecht, The Netherlands}
\affil{$^6$Space Sciences Laboratory, University of California, Berkeley, 
CA 94720-7450}
\affil{$^7$Lawrence Livermore National Laboratory, Livermore, CA 94550}
\affil{$^8$Instituto Astrofisica Spaziale, Rome, Italy}
\affil{$^9$Fermi National Accelerator Laboratory, Batavia, IL 60510}

\begin{abstract}

	The Robotic Optical Transient Search Experiment (ROTSE) seeks to 
measure simultaneous and early afterglow optical emission from gamma-ray 
bursts (GRBs).  A search for optical counterparts to six GRBs with 
localization errors of 1 square degree or better produced no detections.  
The earliest limiting sensitivity is $m_{ROTSE} > 13.1$ at 10.85 seconds 
(5 second exposure) after the gamma-ray rise, and the best limit is
$m_{ROTSE} > 16.0$ at 62 minutes (897 second exposure).  These are the most 
stringent limits obtained for GRB optical counterpart brightness in the 
first hour after the burst.  Consideration of the gamma-ray fluence and 
peak flux for these bursts and for GRB990123 indicates that there is not 
a strong positive correlation between optical flux and gamma-ray emission.

\end{abstract}


\section{Introduction}

	The gamma-ray emission of GRBs typically has a duration of the order 
of tens of seconds or less and exhibits little pattern in its very 
pronounced temporal variation.  Until 1997, the brevity of bursts prevented 
multi-wavelength observation which would allow an accurate localization of 
the source.  As a result, the burst mechanism, environment, location and 
energy scale have been elusive.  Since 1997, multi-wavelength afterglow 
observations (eg. \cite{costa97}, \cite{vanpar97}) have established the 
distance to several bursts (eg. \cite{metzger97}, \cite{kulkarni98}) and 
illuminated the physics processes occurring a few hours to days 
after the gamma-ray onset.  The burst mechanism, however, remains a mystery.

	Prompt radiation provides critical detail about the processes of 
the burst itself.  Detection of such emission, in the optical for 
instance, requires instruments with wide field-of-view, rapid response 
and automated operation.  The ROTSE-I CCD telephoto array meets these 
objectives (see \cite{kehoe99}).  Preliminary 
data from the BATSE detectors (\cite{batse}) on-board the Compton 
Gamma-Ray Observatory are used by the GRB Coordinates Network 
(GCN, \cite{gcn}, \cite{bacodine}) to generate triggers about once per 
day providing rough coordinates ($\Delta\theta \sim 10^\circ$) within 
$\sim 5$ seconds of a burst.  A dynamic response points ROTSE-I at 
the trigger coordinates within 3 seconds of their receipt.  The 
discovery of prompt optical emission from GRB990123 (\cite{rotse99}) 
has illustrated properties of early shock development and the 
immediate environment of bursts.  This paper presents a further search 
for optical counterparts in a subset of our GRB trigger data, as well as
a comparison of the results with GRB990123.

\section{Observations and Reduction}

	The subset of 6 triggers discussed here were taken in the first 
year of operation.  They were selected for this analysis because they
possess localization errors of about 1 square degree or smaller 
(see Table \ref{tab:sixgrbs}).  This positional accuracy, which reduces 
the search area and background by a factor of more than 200 from that 
available from BATSE alone, is generally obtained from 
the relative timing of signals from BATSE and the gamma-ray detectors 
on-board Ulysses (\cite{ulysses}).  Thin 'Interplanetary Network' (IPN) 
annuli are generated which are only about $0.1^\circ$ wide (\cite{ipn}).  
The intersection of the BATSE position probability distribution 
with such a timing annulus produces an IPN arc a few degrees long.  
If available, a third detection reduces this arc to a smaller 
diamond-shaped region.  In the current sample, 4 bursts are localized 
to IPN arcs, and GRB981121 has an IPN diamond using NEAR data.  
The BeppoSAX (\cite{feroci97}, \cite{jager97}) satellite observed 
GRB980329 (\cite{329a_xray}), providing a very accurate position.
This sample does not include GRBs on the faint end of the BATSE fluence 
distribution or short bursts (see \cite{shorthard93}, 
\cite{shorthard95}).  Both limitations will be addressed in later analyses.

	For prompt GRB triggers, we initially begin taking 5 second exposures 
to retain sensitivity to rapid variation, then lengthen to 25 second and 
125 second exposures to maximize sensitivity.  If the trigger position 
error is of the same order as the ROTSE-I field-of-view ($16^\circ \times 
16^\circ$), 
we also 'tile' around the given position at specific epochs in the 
sequence to ensure coverage of sources with errant initial positions but 
well-localized later.  We then return to the direct pointing with longer 
exposures and begin the sequence again.

	ROTSE-I first triggered on GRB980329 and began the first
exposures 11.5 seconds after the burst had started.  Unfortunately, the 
sky was cloudy for the early data and hazy for the 
later images.  Nevertheless, some early images are clear enough in the 
immediate region of the burst to detect 10th magnitude objects.  To
maximize sensitivity in later, clearer images, the last two observations 
are the result of co-adding two and three frames, respectively.
GRB980401 occurred during focusing tests so a manual response of eight 
exposures was performed.  The last four 125 second exposures were 
co-added into one observation spread over 897 seconds.  The final 
localization for GRB980420 places it near the galactic plane (most 
probable $g_b \sim -10^\circ$), and focused images are very crowded.  
The optics of the two cameras covering the main part of the IPN arc, 
however, were poorly focused.  The final localization for GRB980627 
places a majority of the probable area outside of even tiled exposures with
the result that we have 40\% coverage in only four tiled images.  Observing 
conditions for GRB981121 and GRB981223 were good.

	Raw images are dark subtracted and flat-fielded, followed by 
source finding in the corrected images using SExtractor (\cite{sextractor}).  
We then perform an astrometric and photometric calibration by comparison 
to the Hipparcos catalog (\cite{tychocat}).  Our astrometric errors are 
1.4 arcsec.  Since we operate with unfiltered CCDs to maximize 
light-gathering ability, photometry is established by comparing raw ROTSE 
magnitudes to V-band measures and color correcting based on $B-V$.  The 
resulting magnitude, $m_{ROTSE}$, corresponds on average to $m_V$ but 
includes sensitivity in the B, V, I and especially R bands.  Our photometric 
errors are 0.02 magnitude for stars brighter than magnitude 12.

\section{Analysis and Discussion}

	Due to observation of an X-ray counterpart for GRB980329, optical 
(\cite{gcnc41}, \cite{gcnc48}, \cite{gcnc52}) 
and radio (\cite{329a_radio}) counterparts were observed several hours 
later.  For such precise localizations, we would accept any detection 
at the known location of the burst.  No optical emission was observed, 
so for the early images we take the sensitivity to be 0.5 
magnitudes brighter than the dimmest SAO and GSC stars visible in the 
immediate region of the burst.  We calculated the limiting sensitivities 
for the co-added frames by extrapolation of the Hipparcos-derived 
calibration to our 
$5\sigma$ threshold.  A cross-check of this calibration was performed by 
directly comparing to the USNO catalog and finding the faintest $m_{ROTSE}$ 
to which we are more than 50\% efficient.  Given the afterglow 
measurements, we are able to place a constraint on the overall power-law 
decline of optical emission from GRB980329 to be shallower than 
$t^{-1.8}$ with respect to the earliest afterglow detection.  This contrasts
with the faster decline of the X-ray emission (\cite{gcnc59}).

	For the other bursts, we require that a source magnitude vary by 
at least 
$0.5 + 5\sigma$ where $\sigma$ is the statistical error on the dimmest 
measurement.  Varying objects are only considered bona fide optical 
counterpart candidates if they appear in at least two successive 
images.  This removes backgrounds such as cosmic rays and satellite 
glints which show up frequently in ROTSE-I images.  No rapidly varying
objects were found in the allowed error regions of these five bursts.
The image sensitivities were determined from the Hipparcos calibration
as described above, and in most cases were cross-checked with the USNO
catalog comparison.

	Limiting magnitudes are given in Table \ref{tab:sixlims} for 
up to three epochs marking significant improvements in sensitivity.  
Figure \ref{fig:grb_timehist} displays all relevant observations where 
coverage exceeded 50\%, and indicates that ROTSE-I has sensitivity to 
optical bursts significantly fainter than GRB990123.  The earliest limit 
is $m_{ROTSE} > 13.1$ at 10.85 seconds for GRB981223.  The best limit of 
this sample is $ m_{ROTSE} > 16.0$ at 62 minutes for GRB980401.  We can 
conclude that bright optical counterparts (ie. $m_{ROTSE} \sim 10$) are 
uncommon. 

	Since prompt optical emission has been seen in GRB990123, we 
ask whether optical emission from a GRB is correlated with gamma-ray 
output, as is suggested by \cite{sari99}.  Because we do not know whether 
fluence or peak flux are accurate measures of the total gamma ray emission, 
we consider both in our comparison.  To bring all bursts onto a common 
footing, we first adjust their $m_{ROTSE}$ limits by 
$2.5\log(f/f_{GRB990123})$, where $f$ is the gamma-ray fluence.  We 
calculate this fluence to be that measured in the BATSE 50 - 100 keV plus 
100 - 300 keV channels to avoid systematics due to problems in spectral 
fitting the other channels (\cite{mbriggs}).  These fluence-scaled limits 
are plotted along with the GRB990123 observations in Figure 
\ref{fig:rfluence}.  We have also adjusted our optical limits by scaling 
according to the BATSE measure of peak flux in the 64ms binning of the 
50 - 300 keV data (see Figure \ref{fig:rflux}).

	Variation of galactic extinction over the IPN arcs prevents us 
from quoting an accurate value for most of these bursts.  However, it
is much less than 1 magnitude at their most probable locations.  Since 
GRB990123 has a similar low value of extinction (= 0.04), the effect of 
galactic extinction on our comparison should be minimal.  The one 
exception is GRB980420 which may have over 2 magnitudes of extinction.  
Although extinction near the source can only be measured for GRB980329, 
it is likely most GRBs are not so heavily obscured since the great 
majority observed to have both X-ray and radio counterparts also 
exhibit optical emission (\cite{frail99}).

	Under the assumption of gamma-ray scaling, ROTSE-I is 
sensitive to GRB990123-like optical bursts for GRB981223 from 
30 to 300 seconds.  Around 1 minute, the optical emission of 
GRB981121 and GRB981223 would have been more than 2 magnitudes over our 
detection threshold.  Either GRB990123 is atypical of GRBs in 
general, or there is not a strong correlation of optical flux 
with gamma-ray emission and the inherent dispersion 
to any actual correlation must be larger than two magnitudes to 
explain the results from GRB981121 and GRB981223.

\section{Conclusions} 

	In a study of six well-localized gamma-ray bursts, no optical 
counterparts were identified.  When comparing to afterglow observations 
of GRB980329, we constrain the overall power-law decay of the optical 
emission to be shallower than $t^{-1.8}$.  When using either gamma-ray 
fluence or peak flux as a predictor of optical emission, we find
that especially around 1 minute optical emission is at least two 
magnitudes dimmer than for GRB990123.  This non-detection of another 
optical burst indicates that optical emission is not strongly 
correlated with gamma-ray output. 

\begin{acknowledgments}
	We thank the BATSE team for their GRB data and Michael Briggs 
for further assistance.  We also thank the NEAR team for their 
GRB981121 data.  ROTSE is supported by NASA 
under $SR\&T$ grant NAG5-5101, the NSF under grants AST-9703282 and 
AST-9970818, the Research Corporation, the University of Michigan, and the 
Planetary Society.  Work performed at LANL is supported by the DOE under
contract W-7405-ENG-36.  Work performed at LLNL is supported by the DOE
under contract W-7405-ENG-48.
\end{acknowledgments}

%\clearpage

%\clearpage

\figcaption[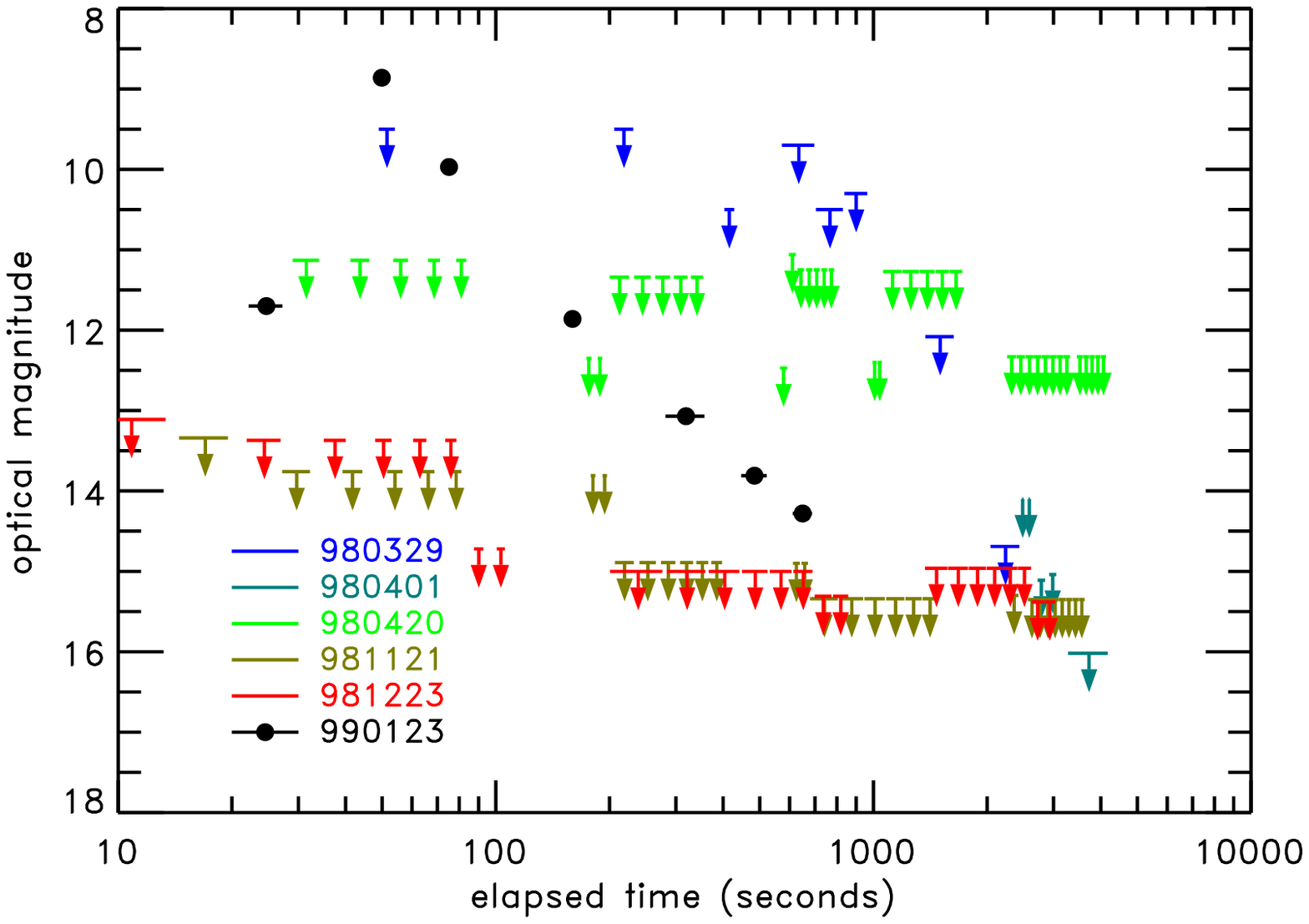]{$m_{ROTSE}$ limiting magnitudes vs. time 
	after gamma-ray onset.  GRB990123 optical burst detections 
	are shown for comparison.  \label{fig:grb_timehist}}

\figcaption[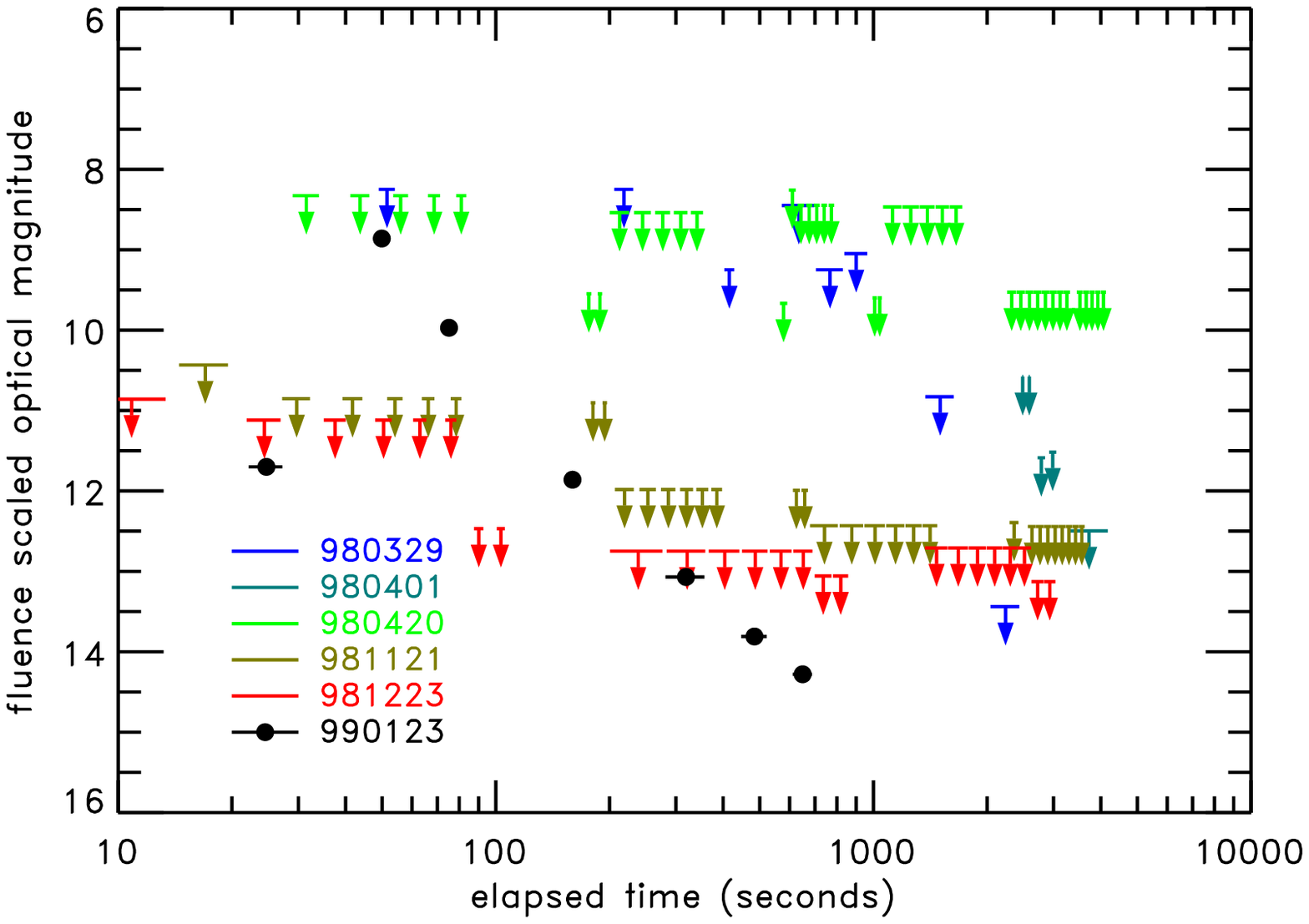]{Limits rescaled by fluence for five 
	GRBs vs. time after gamma-ray onset.  If optical emission were 
	positively correlated with gamma-ray fluence, ROTSE-I would 
	have detected optical bursts for GRB981121 and GRB981223. 
	\label{fig:rfluence}}

\figcaption[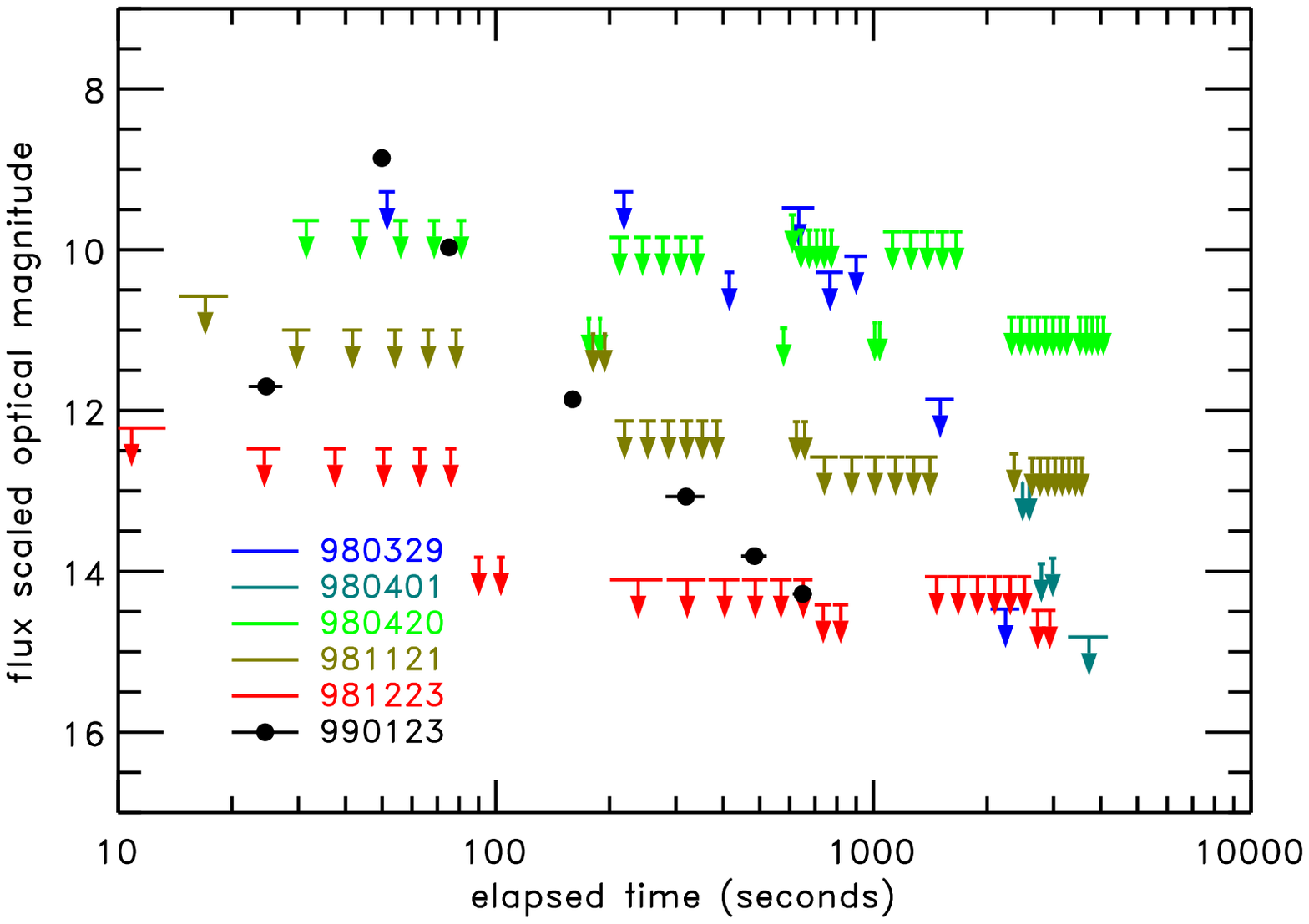]{Flux rescaled limits for five GRBs vs. 
	time after gamma-ray onset.  If optical emission were positively 
	correlated with peak gamma-ray flux, ROTSE-I would have detected 
	optical bursts for GRB981121 and GRB981223. 
	\label{fig:rflux}}

\clearpage

\begin{table}
  \begin{center}
  \begin{tabular}{cccccccc}
	date & trigger & $T_{90}$ & fluence & $\phi_{peak}$ & source & coverage & $t_+$\\
	\hline
	980329	& 6665 &     19   &  322 &      13.8     &    SAX & 100 (100) & 11.5\\
	980401	& 6672 &     28   &   40 &      5.60     &    IPN &       87 & 2478\\
	980420  & 6694 &     40   &   77 &      4.28     &    IPN &  90 (96) & 28.99\\
	980627  & 6880 &     11   &    7 &      1.27     &    IPN &   - (40) & 11.97\\
	981121  & 7219 &     54   &   70 &      1.33     &    IPN & 100 (100) & 14.51\\
	981223  & 7277 &     30   &  128 &      7.44     &    IPN & 100 (67) & 8.35\\
	\hline
	990123  & 7343 &     63   & 1020 &      17.0     &    SAX & 100 (100) & 22.18\\
  \end{tabular}
  \caption{Characteristics of six bursts responded to by ROTSE-I.  For
	comparison, corresponding information for GRB990123 is also 
	given (\cite{rotse99}).  The columns specify: GRB date, BATSE 
	trigger number, duration in seconds, fluence ($\times 10^{-7} 
	erg/cm^2$), peak flux ($\phi_{peak}$, in $photons/cm^2/s$), 
	source of best gamma-ray localization (see text), coverage of 
	the GRB probability (\%), and start time ($t_+$, in sec.) for 
	first image recorded.  Coverages for tiled epochs are indicated 
	in parentheses.
	\label{tab:sixgrbs}}
  \end{center}
\end{table}

\begin{table}
  \begin{center} 
  \begin{tabular}{c|ccc|ccc|ccc}
	date & $t_1$ & $\Delta t_1$ & $m_{ROTSE}(t_1)$ & 
	       $t_2$ & $\Delta t_2$ & $m_{ROTSE}(t_2)$ & 
	       $t_3$ & $\Delta t_3$ & $m_{ROTSE}(t_3)$\\
	\hline
	980329	& 51.5  &   5 &  9.5 &  416    & 25 & 10.5 & 2239 & 390 & 14.7\\
	980401	&    -  &   - &    - & 2485    & 15 & 14.1 & 3726 & 897 & 16.0\\
	980420  & 31.49 &   5 & 11.1 &  176.32 &  5 & 12.4 & 578.21 & 25 & 12.5\\
	980627  &    -  &   - &    - &  180.49 &  5 & 13.2 & 601.87 & 25 & 13.6\\
	981121  & 17.01 &   5 & 13.3 &  219.28 & 25 & 14.9 & 742.05 & 125 & 15.3\\
	981223  & 10.85 &   5 & 13.1 &   90.13 &  5 & 14.7 & 736.43 & 75 & 15.3\\
  \end{tabular}
  \caption{Summary of limits for six bursts responded to by ROTSE-I.  Columns 
	list up to three epochs (middle of exposure, in sec.), and their
	exposure length (sec.) and sensitivity.
	\label{tab:sixlims}}
  \end{center} 
\end{table}

\end{document}